# The Charge-Magnet Paradoxes of Classical Electrodynamics

Masud Mansuripur
College of Optical Sciences, The University of Arizona, Tucson

**Abstract**. A number of charge-magnet paradoxes have been discussed in the literature, beginning with Shockley's famous 1967 paper, where he introduced the notion of hidden momentum in electromagnetic systems. We discuss all these paradoxes in a single, general context, showing that the conservation laws of linear and angular momenta can be satisfied *without* the need for hidden entities, provided that the Einstein-Laub laws of force and torque are used in place of the standard Lorentz law. Einstein and Laub published their paper in 1908, but the simplicity of the conventional Lorentz law overshadowed the subtle features of their formulation which, at first sight, appears somewhat complicated. However, that slight complication turns out to lead to subsequent advantages in light of Shockley's discovery of hidden momentum, which occurred more than a decade after Einstein had passed away. In this paper, we show how the Einstein-Laub formalism handles the underlying problems associated with certain paradoxes of classical electrodynamics involving a static distribution of electric charges and a magnet whose magnetization slowly fades away in time. The Einstein-Laub laws of electromagnetic force and torque treat these paradoxes with elegance and without contradicting the existing body of knowledge, which has been confirmed by more than one and a half century of theoretical and experimental investigations.

**1. Introduction**. In classical electrodynamics, it is generally accepted that neither the mechanical nor the electromagnetic (EM) momentum are conserved by themselves; it is the *total* momentum (i.e., EM plus mechanical) that must be conserved. The continuity equation for momentum involves the EM stress-tensor $\overleftrightarrow{\boldsymbol{\mathcal{T}}}$, the EM momentum-density $\boldsymbol{\wp}$, and the EM force-density $\boldsymbol{f}$, as follows:

$$\overleftrightarrow{\boldsymbol{\nabla}} \cdot \overleftrightarrow{\boldsymbol{\mathcal{T}}}(\boldsymbol{r},t) + \partial_t \boldsymbol{\wp}(\boldsymbol{r},t) + \boldsymbol{f}(\boldsymbol{r},t) = 0. \tag{1}$$

The above equation, although intimately tied to Maxwell's macroscopic equations, is *not* a direct consequence of those equations. In their most general form, the macroscopic equations are[1-5]

$$\boldsymbol{\nabla} \cdot \boldsymbol{D}(\boldsymbol{r},t) = \rho_{\text{free}}(\boldsymbol{r},t), \tag{2a}$$

$$\boldsymbol{\nabla} \times \boldsymbol{H}(\boldsymbol{r},t) = \boldsymbol{J}_{\text{free}}(\boldsymbol{r},t) + \partial_t \boldsymbol{D}(\boldsymbol{r},t), \tag{2b}$$

$$\boldsymbol{\nabla} \times \boldsymbol{E}(\boldsymbol{r},t) = -\partial_t \boldsymbol{B}(\boldsymbol{r},t), \tag{2c}$$

$$\boldsymbol{\nabla} \cdot \boldsymbol{B}(\boldsymbol{r},t) = 0. \tag{2d}$$

Here $\boldsymbol{D}(\boldsymbol{r},t) = \varepsilon_0 \boldsymbol{E}(\boldsymbol{r},t) + \boldsymbol{P}(\boldsymbol{r},t)$ and $\boldsymbol{B}(\boldsymbol{r},t) = \mu_0 \boldsymbol{H}(\boldsymbol{r},t) + \boldsymbol{M}(\boldsymbol{r},t)$, where $\varepsilon_0$ and $\mu_0$ are the permittivity and permeability of free space, respectively, while $\boldsymbol{P}$ and $\boldsymbol{M}$ are the polarization and magnetization of material media. In this standard notation, $\boldsymbol{E}$ is the electric field, $\boldsymbol{H}$ is the magnetic field, $\boldsymbol{D}$ is the displacement, and $\boldsymbol{B}$ is the magnetic induction. Note that $\rho_{\text{free}}$, $\boldsymbol{J}_{\text{free}}$, $\boldsymbol{P}$ and $\boldsymbol{M}$, the sources of the EM fields, are assumed to be arbitrary functions of space-time; in other words, no specific constitutive relations are being introduced into the discussion.[6] Therefore, whether the spatial distribution and the temporal evolution of the sources are predetermined or controlled by the EM fields via certain constitutive relations is not going to affect the expressions of EM force, torque, momentum, and angular momentum presented below.

In the Maxwell-Lorentz formulation (hereinafter shortened to "Lorentz formalism") we have[1]

$$\overleftrightarrow{\boldsymbol{\mathcal{T}}}_L(\boldsymbol{r},t) = ½(\varepsilon_0 \boldsymbol{E} \cdot \boldsymbol{E} + \mu_0^{-1} \boldsymbol{B} \cdot \boldsymbol{B})\overleftrightarrow{\mathbf{I}} - \varepsilon_0 \boldsymbol{E}\boldsymbol{E} - \mu_0^{-1} \boldsymbol{B}\boldsymbol{B}, \tag{3}$$

whereas in the Einstein-Laub formulation,[7]

$$\overleftrightarrow{\mathcal{T}}_{EL}(\boldsymbol{r},t) = ½(\varepsilon_0\boldsymbol{E}\cdot\boldsymbol{E} + \mu_0\boldsymbol{H}\cdot\boldsymbol{H})\overleftrightarrow{\mathrm{I}} - \boldsymbol{DE} - \boldsymbol{BH}. \tag{4}$$

Similarly, according to Chu's theory,[8,9]

$$\overleftrightarrow{\mathcal{T}}_{Chu}(\boldsymbol{r},t) = ½(\varepsilon_0\boldsymbol{E}\cdot\boldsymbol{E} + \mu_0\boldsymbol{H}\cdot\boldsymbol{H})\overleftrightarrow{\mathrm{I}} - \varepsilon_0\boldsymbol{EE} - \mu_0\boldsymbol{HH}, \tag{5}$$

while in Minkowski's theory,[9,10]

$$\overleftrightarrow{\mathcal{T}}_{M}(\boldsymbol{r},t) = ½(\boldsymbol{D}\cdot\boldsymbol{E} + \boldsymbol{B}\cdot\boldsymbol{H})\overleftrightarrow{\mathrm{I}} - \boldsymbol{DE} - \boldsymbol{BH}. \tag{6}$$

The Abraham formulation[11] has the same stress-tensor as that of Minkowski (with the caveat pointed out in the endnote), although their force and momentum densities are different.[12]

Each formalism has its own expressions for EM momentum-density and EM force-density. The Lorentz formalism requires the so-called Livens momentum,[9] $\boldsymbol{\wp}_L(\boldsymbol{r},t) = \varepsilon_0\boldsymbol{E}\times\boldsymbol{B}$, the Einstein-Laub, Abraham, and Chu formalisms all use the Abraham momentum-density $\boldsymbol{\wp}_A(\boldsymbol{r},t) = \boldsymbol{E}\times\boldsymbol{H}/c^2$, while in Minkowski's case, the momentum-density is $\boldsymbol{\wp}_M(\boldsymbol{r},t) = \boldsymbol{D}\times\boldsymbol{B}$.

The difference in $\boldsymbol{\wp}$ between the Abraham and Minkowski formulations results in an additional term, namely, $\partial(\boldsymbol{D}\times\boldsymbol{B} - \boldsymbol{E}\times\boldsymbol{H}/c^2)/\partial t$, which must be added to the Minkowski force-density $\boldsymbol{f}_M(\boldsymbol{r},t)$ in order to arrive at the Abraham force-density $\boldsymbol{f}_A(\boldsymbol{r},t)$.[13-15]

In all cases, the EM angular momentum-density is defined as $\boldsymbol{\ell} = \boldsymbol{r}\times\boldsymbol{\wp}$. Conservation of angular momentum is then a direct consequence of momentum continuity expressed by Eq.(1). This may be verified by cross-multiplying $\boldsymbol{r}$ on the left side of Eq.(1). In the case of Lorentz and Chu formulations, one can readily show that $\boldsymbol{r}\times\overleftrightarrow{\boldsymbol{\nabla}}\cdot\overleftrightarrow{\mathcal{T}}(\boldsymbol{r},t) = \overleftrightarrow{\boldsymbol{\nabla}}\cdot\left[\boldsymbol{r}\times\overleftrightarrow{\mathcal{T}}(\boldsymbol{r},t)\right]$, in which case the corresponding torque-density would be $\boldsymbol{\tau}(\boldsymbol{r},t) = \boldsymbol{r}\times\boldsymbol{f}(\boldsymbol{r},t)$.[12] In the case of Einstein-Laub, Minkowski, and Abraham formulations, $\boldsymbol{r}\times\overleftrightarrow{\boldsymbol{\nabla}}\cdot\overleftrightarrow{\mathcal{T}}(\boldsymbol{r},t) = \overleftrightarrow{\boldsymbol{\nabla}}\cdot\left[\boldsymbol{r}\times\overleftrightarrow{\mathcal{T}}(\boldsymbol{r},t)\right] + \boldsymbol{P}\times\boldsymbol{E} + \boldsymbol{M}\times\boldsymbol{H}$, in which case the extra terms must be bundled with $\boldsymbol{r}\times\boldsymbol{f}$ into an expression for torque-density, yielding $\boldsymbol{\tau}(\boldsymbol{r},t) = \boldsymbol{r}\times\boldsymbol{f}(\boldsymbol{r},t) + \boldsymbol{P}\times\boldsymbol{E} + \boldsymbol{M}\times\boldsymbol{H}$.[12] Clearly, torque is the time-rate-of-exchange of angular momentum between the fields and the material media, just as force is the time-rate-of-exchange of linear momentum.

In the Lorentz formalism, the force and torque densities exerted by EM fields on material media are straightforwardly obtained from Eqs.(1) and (3), as follows:[1-5]

$$\boldsymbol{f}_L(\boldsymbol{r},t) = (\rho_{\text{free}} - \boldsymbol{\nabla}\cdot\boldsymbol{P})\boldsymbol{E} + (\boldsymbol{J}_{\text{free}} + \partial_t\boldsymbol{P} + \mu_0^{-1}\boldsymbol{\nabla}\times\boldsymbol{M})\times\boldsymbol{B}, \tag{7a}$$

$$\boldsymbol{\tau}_L(\boldsymbol{r},t) = \boldsymbol{r}\times\boldsymbol{f}_L(\boldsymbol{r},t). \tag{7b}$$

The corresponding entities in the Einstein-Laub formalism are found to be[7]

$$\boldsymbol{f}_{EL}(\boldsymbol{r},t) = \rho_{\text{free}}\boldsymbol{E} + \boldsymbol{J}_{\text{free}}\times\mu_0\boldsymbol{H} + (\boldsymbol{P}\cdot\boldsymbol{\nabla})\boldsymbol{E} + \partial_t\boldsymbol{P}\times\mu_0\boldsymbol{H} + (\boldsymbol{M}\cdot\boldsymbol{\nabla})\boldsymbol{H} - \partial_t\boldsymbol{M}\times\varepsilon_0\boldsymbol{E}, \tag{8a}$$

$$\boldsymbol{\tau}_{EL}(\boldsymbol{r},t) = \boldsymbol{r}\times\boldsymbol{f}_{EL}(\boldsymbol{r},t) + \boldsymbol{P}\times\boldsymbol{E} + \boldsymbol{M}\times\boldsymbol{H}. \tag{8b}$$

Similarly, in the Chu formulation, the force-density and torque-density expressions are[8,9]

$$\boldsymbol{f}_{Chu}(\boldsymbol{r},t) = (\rho_{\text{free}} - \boldsymbol{\nabla}\cdot\boldsymbol{P})\boldsymbol{E} + (\boldsymbol{J}_{\text{free}} + \partial_t\boldsymbol{P})\times\mu_0\boldsymbol{H} - (\boldsymbol{\nabla}\cdot\boldsymbol{M})\boldsymbol{H} - \partial_t\boldsymbol{M}\times\varepsilon_0\boldsymbol{E}, \tag{9a}$$

$$\boldsymbol{\tau}_{Chu}(\boldsymbol{r},t) = \boldsymbol{r}\times\boldsymbol{f}_{Chu}(\boldsymbol{r},t). \tag{9b}$$

In Abraham's formulation, the force and torque densities turn out to be somewhat different from those in the Einstein-Laub theory, namely,[12]

$$\boldsymbol{f}_A(\boldsymbol{r},t) = \boldsymbol{f}_{EL}(\boldsymbol{r},t) - \tfrac{1}{2}\boldsymbol{\nabla}(\boldsymbol{P}\cdot\boldsymbol{E} + \boldsymbol{M}\cdot\boldsymbol{H}), \tag{10a}$$

$$\boldsymbol{\tau}_A(\boldsymbol{r},t) = \boldsymbol{r}\times\boldsymbol{f}_A(\boldsymbol{r},t) + \boldsymbol{P}\times\boldsymbol{E} + \boldsymbol{M}\times\boldsymbol{H}. \tag{10b}$$

Finally, the force and torque densities in Minkowski's formulation are found to be[12]

$$\boldsymbol{f}_M(\boldsymbol{r},t) = \rho_{\text{free}}\boldsymbol{E} + \boldsymbol{J}_{\text{free}}\times\boldsymbol{B} + [(\boldsymbol{P}\cdot\boldsymbol{\nabla})\boldsymbol{E} + \boldsymbol{P}\times(\boldsymbol{\nabla}\times\boldsymbol{E}) - \tfrac{1}{2}\boldsymbol{\nabla}(\boldsymbol{P}\cdot\boldsymbol{E})]$$
$$+[(\boldsymbol{M}\cdot\boldsymbol{\nabla})\boldsymbol{H} + \boldsymbol{M}\times(\boldsymbol{\nabla}\times\boldsymbol{H}) - \tfrac{1}{2}\boldsymbol{\nabla}(\boldsymbol{M}\cdot\boldsymbol{H})], \tag{11a}$$

$$\boldsymbol{\tau}_M(\boldsymbol{r},t) = \boldsymbol{r}\times\boldsymbol{f}_M(\boldsymbol{r},t) + \boldsymbol{P}\times\boldsymbol{E} + \boldsymbol{M}\times\boldsymbol{H}. \tag{11b}$$

In the special case of linear, isotropic, lossless, non-dispersive media, where the operative constitutive relations are $\boldsymbol{P}(\boldsymbol{r},t) = \varepsilon_0[\varepsilon(\boldsymbol{r}) - 1]\boldsymbol{E}(\boldsymbol{r},t)$ and $\boldsymbol{M}(\boldsymbol{r},t) = \mu_0[\mu(\boldsymbol{r}) - 1]\boldsymbol{H}(\boldsymbol{r},t)$, the Minkowski expressions for force and torque densities may be simplified as follows:[9,12,16]

$$\boldsymbol{f}_M(\boldsymbol{r},t) = \rho_{\text{free}}\boldsymbol{E} + \boldsymbol{J}_{\text{free}}\times\boldsymbol{B} - \tfrac{1}{2}\varepsilon_0(\boldsymbol{\nabla}\varepsilon)(\boldsymbol{E}\cdot\boldsymbol{E}) - \tfrac{1}{2}\mu_0(\boldsymbol{\nabla}\mu)(\boldsymbol{H}\cdot\boldsymbol{H}). \tag{12a}$$

$$\boldsymbol{\tau}_M(\boldsymbol{r},t) = \boldsymbol{r}\times\boldsymbol{f}_M(\boldsymbol{r},t). \tag{12b}$$

In the above equations, the relative permittivity $\varepsilon(\boldsymbol{r})$ and the relative permeability $\mu(\boldsymbol{r})$ are real-valued functions of the spatial coordinates. (Lossless media have real-valued $\varepsilon$ and $\mu$, since loss and gain are associated with the imaginary parts of these parameters.[1,4,5] The $\varepsilon$ and $\mu$ of dispersive media are functions of the frequency of the exciting $\boldsymbol{E}$ and $\boldsymbol{H}$ fields, whereas $\boldsymbol{P}$ and $\boldsymbol{M}$ of non-dispersive media are directly proportional to $\boldsymbol{E}$ and $\boldsymbol{H}$, respectively, regardless of the temporal behavior of these fields.)

In this paper, we will not be concerned with Chu, Minkowski, and Abraham formulations, although their treatment should parallel those of the Lorentz and Einstein-Laub formalisms discussed in the following sections.

**2. Charge-magnet system in the Einstein-Laub formalism**. A typical charge-magnet paradox involves a stationary charge distribution $\rho(\boldsymbol{r})$ and a stationary magnet whose magnetization $\boldsymbol{M}(\boldsymbol{r},t)$ slowly varies as a function of time. These problems are usually analyzed in the Lorentz formalism, where questions often arise with regard to the consistency of classical electrodynamics with the conservation laws and with the special theory of relativity.[17-47] The goal of the present section is to analyze the general charge-magnet problem in the Einstein-Laub formalism, where force and torque as well as linear and angular momenta can be straightforwardly calculated and shown to satisfy the requirements of the conservation laws. In Section 3 we will repeat the same analysis in the Lorentz formalism, where clear differences with the results of the present section will emerge. Section 4 will then introduce the so-called "hidden momentum" into the Lorentz formalism, which succeeds in eliminating the discrepancies.

Let us first demonstrate that a magnet, sitting by itself in field-free vacuum and specified by its static magnetization profile $\boldsymbol{M}(\boldsymbol{r})$, does not move, nor does it rotate under the influence of its own magnetic field. In the Einstein-Laub formalism, the self-force on the magnet is given by

$$\boldsymbol{F}_{EL} = \iiint_{-\infty}^{\infty}(\boldsymbol{M}\cdot\boldsymbol{\nabla})\boldsymbol{H}dv = \iiint_{-\infty}^{\infty}\left[\partial_x(M_x\boldsymbol{H}) + \partial_y\left(M_y\boldsymbol{H}\right) + \partial_z(M_z\boldsymbol{H}) - (\boldsymbol{\nabla}\cdot\boldsymbol{M})\boldsymbol{H}\right]dv$$
$$= -\iiint_{-\infty}^{\infty}(\boldsymbol{\nabla}\cdot\boldsymbol{M})\boldsymbol{H}dv = \iiint_{-\infty}^{\infty}[\boldsymbol{\nabla}\cdot\boldsymbol{M}(\boldsymbol{r})]\iiint_{-\infty}^{\infty}\frac{[\boldsymbol{\nabla}'\cdot\boldsymbol{M}(\boldsymbol{r}')](\boldsymbol{r}-\boldsymbol{r}')}{4\pi\mu_0|\boldsymbol{r}-\boldsymbol{r}'|^3}dv'dv$$
$$= \frac{1}{4\pi\mu_0}\iiint\iiint_{-\infty}^{\infty}[\boldsymbol{\nabla}\cdot\boldsymbol{M}(\boldsymbol{r})][\boldsymbol{\nabla}'\cdot\boldsymbol{M}(\boldsymbol{r}')]\frac{\boldsymbol{r}-\boldsymbol{r}'}{|\boldsymbol{r}-\boldsymbol{r}'|^3}dv'dv. \tag{13}$$

Here the complete differentials appearing in the integrand in the first line of Eq.(13) have been omitted in subsequent lines, simply because their integrals over the entire space vanish. The domain of integration in Eq. (13) contains all values of $\boldsymbol{r}$ and $\boldsymbol{r}'$ inside the magnet. For every pair of points $\boldsymbol{r}$ and $\boldsymbol{r}'$, exchanging $\boldsymbol{r}$ and $\boldsymbol{r}'$ causes their equal but opposite contributions to the integral to cancel out. The total self-force on the magnet thus vanishes. A similar argument applies to the self-torque, namely,

$$
\begin{aligned}
\boldsymbol{T}_{EL} &= \iiint_{-\infty}^{\infty}[\boldsymbol{r}\times(\boldsymbol{M}\cdot\boldsymbol{\nabla})\boldsymbol{H}+\boldsymbol{M}\times\boldsymbol{H}]dv \\
&= \iiint_{-\infty}^{\infty}\big[\partial_x(M_x\boldsymbol{r}\times\boldsymbol{H})+\partial_y\big(M_y\boldsymbol{r}\times\boldsymbol{H}\big)+\partial_z(M_z\boldsymbol{r}\times\boldsymbol{H})-(\boldsymbol{\nabla}\cdot\boldsymbol{M})(\boldsymbol{r}\times\boldsymbol{H})-\boldsymbol{M}\times\boldsymbol{H}+\boldsymbol{M}\times\boldsymbol{H}\big]dv \\
&= -\iiint_{-\infty}^{\infty}(\boldsymbol{\nabla}\cdot\boldsymbol{M})(\boldsymbol{r}\times\boldsymbol{H})dv = \iiint_{-\infty}^{\infty}[\boldsymbol{\nabla}\cdot\boldsymbol{M}(\boldsymbol{r})]\boldsymbol{r}\times\iiint_{-\infty}^{\infty}\frac{[\boldsymbol{\nabla}'\cdot\boldsymbol{M}(\boldsymbol{r}')](\boldsymbol{r}-\boldsymbol{r}')}{4\pi\mu_0|\boldsymbol{r}-\boldsymbol{r}'|^3}dv'dv \\
&= \frac{1}{4\pi\mu_0}\iiint\iiint_{-\infty}^{\infty}[\boldsymbol{\nabla}\cdot\boldsymbol{M}(\boldsymbol{r})][\boldsymbol{\nabla}'\cdot\boldsymbol{M}(\boldsymbol{r}')]\frac{\boldsymbol{r}'\times\boldsymbol{r}}{|\boldsymbol{r}-\boldsymbol{r}'|^3}dv'dv. \qquad (14)
\end{aligned}
$$

Once again, the complete differentials have been omitted after the second line of the above equation, because they integrate to zero. The self-torque in Eq. (14) now vanishes because, once again, exchanging $\boldsymbol{r}$ and $\boldsymbol{r}'$ will switch the sign of the integrand.

Next, we prove that the total EM (Abraham) momentum $\boldsymbol{\mathcal{P}}_A$ is zero for a constant (but otherwise arbitrary) magnetization distribution $\boldsymbol{M}(\boldsymbol{r})$ in the presence of a static $E$-field produced by a static charge distribution $\rho(\boldsymbol{r})$. Since the $E$-field in this case is derived directly from the gradient of the corresponding scalar potential $\psi(\boldsymbol{r})$, we write

$$
\begin{aligned}
\boldsymbol{\mathcal{P}}_A &= \iiint_{-\infty}^{\infty}c^{-2}\boldsymbol{E}(\boldsymbol{r})\times\boldsymbol{H}(\boldsymbol{r})dv = -c^{-2}\iiint_{-\infty}^{\infty}[\boldsymbol{\nabla}\psi(\boldsymbol{r})]\times\boldsymbol{H}(\boldsymbol{r})dv \\
&= -c^{-2}\iiint_{-\infty}^{\infty}[\boldsymbol{\nabla}\times(\psi\boldsymbol{H})-\psi\cancelto{0}{\boldsymbol{\nabla}\times\boldsymbol{H}}]dv. \qquad (15)
\end{aligned}
$$

The last term on the right-hand-side of Eq. (15) vanishes because the curl of the $H$-field produced by a static magnet is zero. The remaining term is a complete differential, whose integral also vanishes for a finite distribution of $\boldsymbol{M}(\boldsymbol{r})$ and $\rho(\boldsymbol{r})$. (Alternatively, one could invoke the identity $\iiint\boldsymbol{\nabla}\times\boldsymbol{A}\,dv=\oiint\boldsymbol{n}\times\boldsymbol{A}\,ds$ in order to convert the volume integral of $\boldsymbol{\nabla}\times(\psi\boldsymbol{H})$ to a surface integral at infinity.) Therefore, a static magnet in the presence of a static charge distribution has no EM (Abraham) momentum.

**2.1. Conservation of linear momentum as the magnet gradually loses its magnetization**. If the magnet now warms up slowly so that its magnetization profile, while declining, could be considered static at any given time, the following argument reveals that the net force of the induced $E$-field on the charge distribution $\rho(\boldsymbol{r})$ will be equal and opposite to the net force of the $E$-field produced by $\rho(\boldsymbol{r})$ on the (slowly) time-varying magnetization $\boldsymbol{M}(\boldsymbol{r},t)$. We are assuming, of course, that the slowly-warming magnet does not exert a force on itself. This is because, in the Einstein-Laub expression of force-density, the *induced* $E$-field, a small entity, is multiplied into $\partial\boldsymbol{M}(\boldsymbol{r},t)/\partial t$, another small entity. Moreover, for a nearly-static magnet, $(\boldsymbol{M}\cdot\boldsymbol{\nabla})\boldsymbol{H}$ integrates to zero, as can be seen below:

$$
\begin{aligned}
(\boldsymbol{M}\cdot\boldsymbol{\nabla})\boldsymbol{H} &= (\boldsymbol{B}\cdot\boldsymbol{\nabla})\boldsymbol{H}-\mu_0(\boldsymbol{H}\cdot\boldsymbol{\nabla})\boldsymbol{H} \\
&= \partial_x(B_x\boldsymbol{H})+\partial_y\big(B_y\boldsymbol{H}\big)+\partial_z(B_z\boldsymbol{H})-(\cancelto{0}{\boldsymbol{\nabla}\cdot\boldsymbol{B}})\boldsymbol{H}-\boldsymbol{\nabla}(½\mu_0\boldsymbol{H}\cdot\boldsymbol{H})+\mu_0\boldsymbol{H}\times(\cancelto{0}{\boldsymbol{\nabla}\times\boldsymbol{H}}). \qquad (16)
\end{aligned}
$$

Since, in general, $\boldsymbol{\nabla}\cdot\boldsymbol{B}=0$ and, for a static magnet, $\boldsymbol{\nabla}\times\boldsymbol{H}=0$, and also since the remaining terms in Eq.(16) are complete differentials, the integral of $(\boldsymbol{M}\cdot\boldsymbol{\nabla})\boldsymbol{H}$ over the volume of the magnet must vanish. [The same result was obtained earlier, albeit from a different perspective, in the discussion surrounding Eq.(13).] The bottom line is that the magnet does *not* experience a net self-force, whether its magnetization is static or has a slow time dependence.

For a piece of magnetic material, whose magnetization $\boldsymbol{M}(\boldsymbol{r})$ slowly vanishes in the presence of a stationary charge distribution $\rho(\boldsymbol{r})$, the time-rate-of-change of the EM (Abraham) linear momentum is given by

$$\begin{aligned}
\partial_t\boldsymbol{\mathcal{P}}_A &= \frac{\partial}{\partial t}\iiint_{-\infty}^{\infty}\varepsilon_0\boldsymbol{E}(\boldsymbol{r},t)\times\mu_0\boldsymbol{H}(\boldsymbol{r},t)dv\\
&= \iiint_{-\infty}^{\infty}\varepsilon_0\frac{\partial\boldsymbol{E}(\boldsymbol{r},t)}{\partial t}\times\mu_0\boldsymbol{H}(\boldsymbol{r},t)dv+\iiint_{-\infty}^{\infty}\varepsilon_0\boldsymbol{E}(\boldsymbol{r},t)\times\frac{\partial[\boldsymbol{B}(\boldsymbol{r},t)-\boldsymbol{M}(\boldsymbol{r},t)]}{\partial t}dv\\
&= \mu_0\iiint_{-\infty}^{\infty}(\boldsymbol{\nabla}\times\boldsymbol{H})\times\boldsymbol{H}dv-\varepsilon_0\iiint_{-\infty}^{\infty}\boldsymbol{E}\times(\boldsymbol{\nabla}\times\boldsymbol{E})dv+\iiint_{-\infty}^{\infty}[\partial_t\boldsymbol{M}(\boldsymbol{r},t)]\times\varepsilon_0\boldsymbol{E}dv\\
&= -\iiint_{-\infty}^{\infty}[\boldsymbol{\nabla}(\tfrac{1}{2}\mu_0\boldsymbol{H}\cdot\boldsymbol{H}+\tfrac{1}{2}\varepsilon_0\boldsymbol{E}\cdot\boldsymbol{E})-\mu_0(\boldsymbol{H}\cdot\boldsymbol{\nabla})\boldsymbol{H}-\varepsilon_0(\boldsymbol{E}\cdot\boldsymbol{\nabla})\boldsymbol{E}]dv+\iiint_{-\infty}^{\infty}(\partial_t\boldsymbol{M})\times\varepsilon_0\boldsymbol{E}dv\\
&= -\iiint_{-\infty}^{\infty}[\boldsymbol{\nabla}(\tfrac{1}{2}\mu_0\boldsymbol{H}\cdot\boldsymbol{H}+\tfrac{1}{2}\varepsilon_0\boldsymbol{E}\cdot\boldsymbol{E})-(\boldsymbol{B}\cdot\boldsymbol{\nabla})\boldsymbol{H}-\varepsilon_0(\boldsymbol{E}\cdot\boldsymbol{\nabla})\boldsymbol{E}]dv\\
&\quad -\iiint_{-\infty}^{\infty}[(\boldsymbol{M}\cdot\boldsymbol{\nabla})\boldsymbol{H}-(\partial_t\boldsymbol{M})\times\varepsilon_0\boldsymbol{E}]dv. \qquad (17)
\end{aligned}$$

Now, in the preceding equation,

$$(\boldsymbol{B}\cdot\boldsymbol{\nabla})\boldsymbol{H}=\partial_x(B_x\boldsymbol{H})+\partial_y(B_y\boldsymbol{H})+\partial_z(B_z\boldsymbol{H})-(\boldsymbol{\nabla}\cdot\boldsymbol{B})\boldsymbol{H}. \qquad (18)$$

$$\varepsilon_0(\boldsymbol{E}\cdot\boldsymbol{\nabla})\boldsymbol{E}=\varepsilon_0[\partial_x(E_x\boldsymbol{E})+\partial_y(E_y\boldsymbol{E})+\partial_z(E_z\boldsymbol{E})]-\rho(\boldsymbol{r})\boldsymbol{E}. \qquad (19)$$

Taking note of the fact that, in Eq.(17), the complete differentials integrate to zero, we will have

$$\partial_t\boldsymbol{\mathcal{P}}_A=-\iiint_{-\infty}^{\infty}[\rho(\boldsymbol{r})\boldsymbol{E}+(\boldsymbol{M}\cdot\boldsymbol{\nabla})\boldsymbol{H}-(\partial_t\boldsymbol{M})\times\varepsilon_0\boldsymbol{E}]dv. \qquad (20)$$

The leading minus-sign aside, the right-hand-side of Eq.(20) is the instantaneous force exerted on the charge distribution (1st term) plus the force exerted on the magnet (2nd and 3rd terms). Since, as discussed earlier, the self-forces are negligible, the total force on $\rho(\boldsymbol{r})$ arises from the induced $E$-field of the magnet, while the total force on the magnet arises from the $E$-field produced by $\rho(\boldsymbol{r})$. Considering that the left-hand-side of Eq.(20) is negligibly small at all times [see the discussion surrounding Eq.(15)], the aforesaid forces must be equal and opposite to each other. The bottom line is that the mechanical linear momentum imparted to the electric charges by the slowly fading magnetization is equal in magnitude and opposite in direction to the mechanical linear momentum imparted to the magnet by the electric charges.

**2.2. Electromagnetic angular momentum of the static charge-magnet system**. In the present section and the next, we address the conservation of angular momentum in the Einstein-Laub formalism. Unlike the linear momentum, the EM (Abraham) angular momentum of the static charge-magnet system is not necessarily zero, as can be seen from the following calculation:

$$\begin{aligned}
\boldsymbol{\mathcal{L}}_A &= \iiint_{-\infty}^{\infty}\boldsymbol{r}\times(\varepsilon_0\boldsymbol{E}\times\mu_0\boldsymbol{H})dv=\iiint_{-\infty}^{\infty}\boldsymbol{r}\times[\varepsilon_0\boldsymbol{E}\times(\boldsymbol{B}-\boldsymbol{M})]dv\\
&= \iiint_{-\infty}^{\infty}\boldsymbol{r}\times(\boldsymbol{M}\times\varepsilon_0\boldsymbol{E})dv+\iiint_{-\infty}^{\infty}\boldsymbol{r}\times[\varepsilon_0\boldsymbol{E}\times(\boldsymbol{\nabla}\times\boldsymbol{A})]dv. \qquad (21)
\end{aligned}$$

Here we have introduced the vector potential $\boldsymbol{A}(\boldsymbol{r})$ produced by the magnetization $\boldsymbol{M}(\boldsymbol{r})$. The second integrand on the right-hand-side of Eq.(21) may be simplified as follows:

$$\begin{aligned}
\boldsymbol{r}\times[\varepsilon_0\boldsymbol{E}\times(\boldsymbol{\nabla}\times\boldsymbol{A})] &= \boldsymbol{r}\times\varepsilon_0[\boldsymbol{\nabla}(\boldsymbol{E}\cdot\boldsymbol{A})-(\boldsymbol{E}\cdot\boldsymbol{\nabla})\boldsymbol{A}-(\boldsymbol{A}\cdot\boldsymbol{\nabla})\boldsymbol{E}-\boldsymbol{A}\times\overset{0}{(\boldsymbol{\nabla}\times\boldsymbol{E})}]\\
&= \boldsymbol{r}\times\varepsilon_0\{\boldsymbol{\nabla}(\boldsymbol{E}\cdot\boldsymbol{A})-[\partial_x(E_x\boldsymbol{A})+\partial_y(E_y\boldsymbol{A})+\partial_z(E_z\boldsymbol{A})-(\boldsymbol{\nabla}\cdot\boldsymbol{E})\boldsymbol{A}]\\
&\qquad -[\partial_x(A_x\boldsymbol{E})+\partial_y(A_y\boldsymbol{E})+\partial_z(A_z\boldsymbol{E})-\overset{0}{(\boldsymbol{\nabla}\cdot\boldsymbol{A})}\boldsymbol{E}]\}\\
&= \varepsilon_0\boldsymbol{r}\times\boldsymbol{\nabla}(\boldsymbol{E}\cdot\boldsymbol{A})-\varepsilon_0\boldsymbol{r}\times[\partial_x(E_x\boldsymbol{A})+\partial_y(E_y\boldsymbol{A})+\partial_z(E_z\boldsymbol{A})]\\
&\quad +\boldsymbol{r}\times\rho(\boldsymbol{r})\boldsymbol{A}-\varepsilon_0\boldsymbol{r}\times[\partial_x(A_x\boldsymbol{E})+\partial_y(A_y\boldsymbol{E})+\partial_z(A_z\boldsymbol{E})]\\
&= \varepsilon_0\{(\boldsymbol{E}\cdot\boldsymbol{A})\overset{0}{(\boldsymbol{\nabla}\times\boldsymbol{r})}-\boldsymbol{\nabla}\times[(\boldsymbol{E}\cdot\boldsymbol{A})\boldsymbol{r}]\}\\
&\quad -\varepsilon_0[\partial_x(\boldsymbol{r}\times E_x\boldsymbol{A})+\partial_y(\boldsymbol{r}\times E_y\boldsymbol{A})+\partial_z(\boldsymbol{r}\times E_z\boldsymbol{A})-\boldsymbol{E}\times\boldsymbol{A}]+\boldsymbol{r}\times\rho(\boldsymbol{r})\boldsymbol{A}\\
&\quad -\varepsilon_0[\partial_x(\boldsymbol{r}\times A_x\boldsymbol{E})+\partial_y(\boldsymbol{r}\times A_y\boldsymbol{E})+\partial_z(\boldsymbol{r}\times A_z\boldsymbol{E})-\boldsymbol{A}\times\boldsymbol{E}]. \qquad (22)
\end{aligned}$$

The complete differentials in the preceding expression integrate to zero, and we are left with

$$\boldsymbol{\mathcal{L}}_A=\iiint_{-\infty}^{\infty}\boldsymbol{r}\times(\boldsymbol{M}\times\varepsilon_0\boldsymbol{E})dv+\iiint_{-\infty}^{\infty}\boldsymbol{r}\times\rho(\boldsymbol{r})\boldsymbol{A}dv. \qquad (23)$$

To evaluate the right-hand-side of Eq.(23), one must specify $\rho(\boldsymbol{r})$ and $\boldsymbol{M}(\boldsymbol{r})$. As a simple example, consider a point-charge–point-dipole system, with the magnetic dipole $\boldsymbol{m}_0$ located at $\boldsymbol{r}_m$ and the point-charge $q$ sitting at $\boldsymbol{r}_q$. Writing $\boldsymbol{M}(\boldsymbol{r})=\boldsymbol{m}_0\delta(\boldsymbol{r}-\boldsymbol{r}_m)$ and $\rho(\boldsymbol{r})=q\delta(\boldsymbol{r}-\boldsymbol{r}_q)$, we will have

$$\boldsymbol{\mathcal{L}}_A=\boldsymbol{r}_m\times[\boldsymbol{m}_0\times\varepsilon_0\boldsymbol{E}(\boldsymbol{r}_m)]+\boldsymbol{r}_q\times q\boldsymbol{A}(\boldsymbol{r}_q). \qquad (24)$$

Now, $q\boldsymbol{A}(\boldsymbol{r}_q)=q\frac{\boldsymbol{m}_0\times(\boldsymbol{r}_q-\boldsymbol{r}_m)}{4\pi|\boldsymbol{r}_q-\boldsymbol{r}_m|^3}=-\boldsymbol{m}_0\times\varepsilon_0\boldsymbol{E}(\boldsymbol{r}_m)$. Consequently,

$$\boldsymbol{\mathcal{L}}_A=(\boldsymbol{r}_q-\boldsymbol{r}_m)\times q\boldsymbol{A}(\boldsymbol{r}_q). \qquad (25)$$

Clearly, there is a net angular momentum in the EM field, provided that $\boldsymbol{A}(\boldsymbol{r}_q)$ is not zero and also not aligned with the line joining the two particles.

**2.3. Conservation of angular momentum as the magnetization fades**. In this section we show that, upon slow warming-up of the magnet, the net torque exerted on the material system, which consists of the charge $\rho(\boldsymbol{r})$ and the magnetization $\boldsymbol{M}(\boldsymbol{r})$, is equal in magnitude and opposite in direction to the time-rate-of-change of the EM (Abraham) angular momentum $\boldsymbol{\mathcal{L}}_A$. We have

$$\begin{aligned}
\partial_t\boldsymbol{\mathcal{L}}_A &= \frac{\partial}{\partial t}\iiint_{-\infty}^{\infty}\boldsymbol{r}\times[\varepsilon_0\boldsymbol{E}(\boldsymbol{r},t)\times\mu_0\boldsymbol{H}(\boldsymbol{r},t)]dv\\
&= \iiint_{-\infty}^{\infty}\boldsymbol{r}\times\left(\frac{\partial\boldsymbol{D}}{\partial t}\times\mu_0\boldsymbol{H}\right)dv+\iiint_{-\infty}^{\infty}\boldsymbol{r}\times\left(\varepsilon_0\boldsymbol{E}\times\frac{\partial\boldsymbol{B}}{\partial t}\right)dv-\iiint_{-\infty}^{\infty}\boldsymbol{r}\times\left(\varepsilon_0\boldsymbol{E}\times\frac{\partial\boldsymbol{M}}{\partial t}\right)dv\\
&= \iiint_{-\infty}^{\infty}\boldsymbol{r}\times[(\boldsymbol{\nabla}\times\boldsymbol{H})\times\mu_0\boldsymbol{H}]dv-\iiint_{-\infty}^{\infty}\boldsymbol{r}\times[\varepsilon_0\boldsymbol{E}\times(\boldsymbol{\nabla}\times\boldsymbol{E})]dv+\iiint_{-\infty}^{\infty}\boldsymbol{r}\times\left(\frac{\partial\boldsymbol{M}}{\partial t}\times\varepsilon_0\boldsymbol{E}\right)dv\\
&= -\iiint_{-\infty}^{\infty}\boldsymbol{r}\times[\boldsymbol{\nabla}(½\mu_0\boldsymbol{H}\cdot\boldsymbol{H}+½\varepsilon_0\boldsymbol{E}\cdot\boldsymbol{E})-\mu_0(\boldsymbol{H}\cdot\boldsymbol{\nabla})\boldsymbol{H}-\varepsilon_0(\boldsymbol{E}\cdot\boldsymbol{\nabla})\boldsymbol{E}]dv\\
&\quad +\iiint_{-\infty}^{\infty}\boldsymbol{r}\times\left(\frac{\partial\boldsymbol{M}}{\partial t}\times\varepsilon_0\boldsymbol{E}\right)dv. \qquad (26)
\end{aligned}$$

Now, in the preceding equation,

$$\boldsymbol{r}\times\boldsymbol{\nabla}(\tfrac{1}{2}\mu_0\boldsymbol{H}\cdot\boldsymbol{H}+\tfrac{1}{2}\varepsilon_0\boldsymbol{E}\cdot\boldsymbol{E}) = -\boldsymbol{\nabla}\times[(\tfrac{1}{2}\mu_0\boldsymbol{H}\cdot\boldsymbol{H}+\tfrac{1}{2}\varepsilon_0\boldsymbol{E}\cdot\boldsymbol{E})\boldsymbol{r}]. \tag{27}$$

$$\begin{aligned}\boldsymbol{r}\times(\mu_0\boldsymbol{H}\cdot\boldsymbol{\nabla})\boldsymbol{H} &= \boldsymbol{r}\times(\boldsymbol{B}\cdot\boldsymbol{\nabla})\boldsymbol{H}-\boldsymbol{r}\times(\boldsymbol{M}\cdot\boldsymbol{\nabla})\boldsymbol{H} = \partial_x(B_x\boldsymbol{r}\times\boldsymbol{H})+\partial_y(B_y\boldsymbol{r}\times\boldsymbol{H})+\partial_z(B_z\boldsymbol{r}\times\boldsymbol{H})\\ &\quad -\cancelto{0}{(\boldsymbol{\nabla}\cdot\boldsymbol{B})}(\boldsymbol{r}\times\boldsymbol{H})-\boldsymbol{B}\times\boldsymbol{H}-\boldsymbol{r}\times(\boldsymbol{M}\cdot\boldsymbol{\nabla})\boldsymbol{H}\\ &= \partial_x(B_x\boldsymbol{r}\times\boldsymbol{H})+\partial_y(B_y\boldsymbol{r}\times\boldsymbol{H})+\partial_z(B_z\boldsymbol{r}\times\boldsymbol{H})-\boldsymbol{M}\times\boldsymbol{H}-\boldsymbol{r}\times(\boldsymbol{M}\cdot\boldsymbol{\nabla})\boldsymbol{H}.\end{aligned} \tag{28}$$

$$\begin{aligned}\boldsymbol{r}\times(\varepsilon_0\boldsymbol{E}\cdot\boldsymbol{\nabla})\boldsymbol{E} &= \partial_x(\varepsilon_0E_x\boldsymbol{r}\times\boldsymbol{E})+\partial_y(\varepsilon_0E_y\boldsymbol{r}\times\boldsymbol{E})+\partial_z(\varepsilon_0E_z\boldsymbol{r}\times\boldsymbol{E})-(\boldsymbol{\nabla}\cdot\varepsilon_0\boldsymbol{E})(\boldsymbol{r}\times\boldsymbol{E})-\cancelto{0}{\varepsilon_0\boldsymbol{E}\times\boldsymbol{E}}\\ &= \partial_x(\varepsilon_0E_x\boldsymbol{r}\times\boldsymbol{E})+\partial_y(\varepsilon_0E_y\boldsymbol{r}\times\boldsymbol{E})+\partial_z(\varepsilon_0E_z\boldsymbol{r}\times\boldsymbol{E})-\boldsymbol{r}\times\rho(\boldsymbol{r})\boldsymbol{E}.\end{aligned} \tag{29}$$

The complete differentials integrate to zero and we are left with

$$\partial_t\boldsymbol{\mathcal{L}}_A = -\iiint_{-\infty}^{\infty}\{\boldsymbol{r}\times[\rho(\boldsymbol{r})\boldsymbol{E}+(\boldsymbol{M}\cdot\boldsymbol{\nabla})\boldsymbol{H}-(\partial_t\boldsymbol{M})\times\varepsilon_0\boldsymbol{E}]+\boldsymbol{M}\times\boldsymbol{H}\}dv. \tag{30}$$

The leading minus-sign aside, the right-hand-side of the above equation is the Einstein-Laub torque exerted on the charge-magnet system as the magnet warms up. One must distinguish between the $E$-field $\boldsymbol{E}_\rho(\boldsymbol{r})$ produced by $\rho(\boldsymbol{r})$ and that produced by the magnet, $\boldsymbol{E}_m(\boldsymbol{r},t)$. The $E$-field of the charge distribution acting on itself does *not* produce a torque; however, the $E$-field induced by the warming magnet can produce a torque on $\rho(\boldsymbol{r})$. Similarly, the magnet's self $E$-field contributes negligibly to $(\partial_t\boldsymbol{M})\times\varepsilon_0\boldsymbol{E}$, because both terms appearing in the cross-product are vanishingly small, whereas the main contribution to this force term comes from the $E$-field originating in $\rho(\boldsymbol{r})$. The self-torque of the $H$-field acting on the magnet in the steady-state situation (i.e., when $\partial_t\boldsymbol{M}=0$) was shown in Eq.(14) to vanish—i.e., the magnet does *not* rotate spontaneously. Therefore, the remaining torques on the magnet produced by the induced $H$-field can be made negligibly small by making the change in $\boldsymbol{M}$ as slowly as desired (the relevant terms are of the second order in $\partial_t\boldsymbol{M}$). The bottom line is that $\boldsymbol{E}_m(\boldsymbol{r},t)$ imparts some angular momentum to the charge distribution $\rho(\boldsymbol{r})$, while $\boldsymbol{E}_\rho(\boldsymbol{r})$ imparts some angular momentum to the declining magnetization $\boldsymbol{M}(\boldsymbol{r},t)$. In accordance with Eq.(30), the total mechanical angular momentum thus imparted to the charge-magnet system comes out of the EM angular momentum that resides initially in the $\boldsymbol{E}$ and $\boldsymbol{H}$ fields.

**3. Charge-magnet system in the Lorentz formalism**. As in the Einstein-Laub case, the self-force on a static magnet in the Lorentz formalism is found to be zero. To see this we write

$$\begin{aligned}\boldsymbol{f}_L(\boldsymbol{r}) &= [\mu_0^{-1}\boldsymbol{\nabla}\times\boldsymbol{M}(\boldsymbol{r})]\times\boldsymbol{B}(\boldsymbol{r}) = \mu_0^{-1}[\boldsymbol{\nabla}\times(\boldsymbol{B}-\mu_0\boldsymbol{H})]\times\boldsymbol{B}\\ &= \mu_0^{-1}(\boldsymbol{\nabla}\times\boldsymbol{B})\times\boldsymbol{B}-\cancelto{0}{(\boldsymbol{\nabla}\times\boldsymbol{H})}\times\boldsymbol{B} = \mu_0^{-1}[(\boldsymbol{B}\cdot\boldsymbol{\nabla})\boldsymbol{B}-\tfrac{1}{2}\boldsymbol{\nabla}(\boldsymbol{B}\cdot\boldsymbol{B})]\\ &= \mu_0^{-1}\big[\partial_x(B_x\boldsymbol{B})+\partial_y(B_y\boldsymbol{B})+\partial_z(B_z\boldsymbol{B})-\cancelto{0}{(\boldsymbol{\nabla}\cdot\boldsymbol{B})}\boldsymbol{B}-\tfrac{1}{2}\boldsymbol{\nabla}(\boldsymbol{B}\cdot\boldsymbol{B})\big].\end{aligned} \tag{31}$$

The surviving terms in the preceding equation are seen to be complete differentials. Therefore, the self-force on the magnet, obtained by integrating the above $\boldsymbol{f}_L$ over its volume, vanishes. An alternative, albeit somewhat lengthier, derivation of the same result which exploits the properties of the vector potential $\boldsymbol{A}(\boldsymbol{r})$ is as follows:

$$\begin{aligned}\boldsymbol{F}_L &= \iiint_{-\infty}^{\infty}[\mu_0^{-1}\boldsymbol{\nabla}\times\boldsymbol{M}(\boldsymbol{r})]\times\boldsymbol{B}(\boldsymbol{r})dv = \mu_0^{-1}\iiint_{-\infty}^{\infty}[\boldsymbol{\nabla}\times\boldsymbol{M}(\boldsymbol{r})]\times[\boldsymbol{\nabla}\times\boldsymbol{A}(\boldsymbol{r})]dv\\ &= \mu_0^{-1}\iiint_{-\infty}^{\infty}[\boldsymbol{\nabla}\times\boldsymbol{M}(\boldsymbol{r})]\times\left[\boldsymbol{\nabla}\times\iiint_{-\infty}^{\infty}\frac{\boldsymbol{\nabla}'\times\boldsymbol{M}(\boldsymbol{r}')}{4\pi|\boldsymbol{r}-\boldsymbol{r}'|}dv'\right]dv\end{aligned}$$

$$= \frac{1}{4\pi\mu_0} \iiint_{-\infty}^{\infty} [\boldsymbol{\nabla} \times \boldsymbol{M}(\boldsymbol{r})] \times \iiint_{-\infty}^{\infty} \boldsymbol{\nabla}(|\boldsymbol{r} - \boldsymbol{r}'|^{-1}) \times [\boldsymbol{\nabla}' \times \boldsymbol{M}(\boldsymbol{r}')] dv' dv$$

$$= -\frac{1}{4\pi\mu_0} \iiint \iiint_{-\infty}^{\infty} [\boldsymbol{\nabla} \times \boldsymbol{M}(\boldsymbol{r})] \times \left\{ \frac{\boldsymbol{r}-\boldsymbol{r}'}{|\boldsymbol{r}-\boldsymbol{r}'|^3} \times [\boldsymbol{\nabla}' \times \boldsymbol{M}(\boldsymbol{r}')] \right\} dv' dv$$

$$= -\frac{1}{4\pi\mu_0} \left\{ \iiint \iiint_{-\infty}^{\infty} \{ [\boldsymbol{\nabla} \times \boldsymbol{M}(\boldsymbol{r})] \cdot [\boldsymbol{\nabla}' \times \boldsymbol{M}(\boldsymbol{r}')] \} \frac{\boldsymbol{r}-\boldsymbol{r}'}{|\boldsymbol{r}-\boldsymbol{r}'|^3} dv dv' \right.$$

$$\left. - \iiint_{-\infty}^{\infty} [\boldsymbol{\nabla}' \times \boldsymbol{M}(\boldsymbol{r}')] \iiint_{-\infty}^{\infty} \frac{[\boldsymbol{\nabla} \times \boldsymbol{M}(\boldsymbol{r})] \cdot (\boldsymbol{r}-\boldsymbol{r}')}{|\boldsymbol{r}-\boldsymbol{r}'|^3} dv dv' \right\}. \quad (32)$$

The first integral on the right-hand-side of Eq.(32) vanishes because it is anti-symmetric, that is, an exchange of $\boldsymbol{r}$ and $\boldsymbol{r}'$ switches the sign of its integrand. As for the second integral, we have

$$\iiint_{-\infty}^{\infty} \frac{[\boldsymbol{\nabla} \times \boldsymbol{M}(\boldsymbol{r})] \cdot (\boldsymbol{r}-\boldsymbol{r}')}{|\boldsymbol{r}-\boldsymbol{r}'|^3} dv = \iiint_{-\infty}^{\infty} \left\{ \boldsymbol{\nabla} \cdot \left[ \frac{\boldsymbol{M}(\boldsymbol{r}) \times (\boldsymbol{r}-\boldsymbol{r}')}{|\boldsymbol{r}-\boldsymbol{r}'|^3} \right] + \boldsymbol{M}(\boldsymbol{r}) \cdot \boldsymbol{\nabla} \times \left( \frac{\boldsymbol{r}-\boldsymbol{r}'}{|\boldsymbol{r}-\boldsymbol{r}'|^3} \right) \right\} dv = 0. \quad (33)$$

The vanishing of the above integral is due to the fact that the divergence integrates to zero, and also $\boldsymbol{\nabla} \times [(\boldsymbol{r} - \boldsymbol{r}')/|\boldsymbol{r} - \boldsymbol{r}'|^3] = 0$. We conclude once again that the self-force $\boldsymbol{F}_L$ on the magnet is equal to zero. Similar arguments may be advanced to confirm the vanishing of the self-torque. For example, starting with Eq.(31), we will have

$$\boldsymbol{\tau}_L = \boldsymbol{r} \times \boldsymbol{f}_L = \mu_0^{-1} \boldsymbol{r} \times \left[ \partial_x (B_x \boldsymbol{B}) + \partial_y \left( B_y \boldsymbol{B} \right) + \partial_z (B_z \boldsymbol{B}) - ½ \boldsymbol{\nabla}(\boldsymbol{B} \cdot \boldsymbol{B}) \right]$$

$$= \mu_0^{-1} \left\{ \partial_x (\boldsymbol{r} \times B_x \boldsymbol{B}) + \partial_y \left( \boldsymbol{r} \times B_y \boldsymbol{B} \right) + \partial_z (\boldsymbol{r} \times B_z \boldsymbol{B}) - \underbrace{\boldsymbol{B} \times \boldsymbol{B}}_{0} \right.$$

$$\left. + ½ \boldsymbol{\nabla} \times [(\boldsymbol{B} \cdot \boldsymbol{B}) \boldsymbol{r}] - ½ (\boldsymbol{B} \cdot \boldsymbol{B}) (\underbrace{\boldsymbol{\nabla} \times \boldsymbol{r}}_{0}) \right\}. \quad (34)$$

The surviving terms in the above equation are seen to be complete differentials. Therefore, the self-torque experienced by the magnet, obtained by integrating $\boldsymbol{\tau}_L$ over its volume, vanishes.

**3.1. Conservation of linear momentum as the magnetization fades**. Consider a piece of magnetic material having an arbitrary magnetization $\boldsymbol{M}(\boldsymbol{r}, t)$, which varies with time arbitrarily. Considering that the EM momentum in the Lorentz formalism is the Livens momentum, the time-rate-of-change of the field's momentum is related to the EM force experienced by the material media, as follows:

$$\partial_t \boldsymbol{\mathcal{P}}_L = \frac{\partial}{\partial t} \iiint_{-\infty}^{\infty} \varepsilon_0 \boldsymbol{E}(\boldsymbol{r}, t) \times \boldsymbol{B}(\boldsymbol{r}, t) dv$$

$$= \iiint_{-\infty}^{\infty} \varepsilon_0 \frac{\partial \boldsymbol{E}(\boldsymbol{r},t)}{\partial t} \times \boldsymbol{B}(\boldsymbol{r}, t) dv + \iiint_{-\infty}^{\infty} \varepsilon_0 \boldsymbol{E}(\boldsymbol{r}, t) \times \frac{\partial \boldsymbol{B}(\boldsymbol{r},t)}{\partial t} dv$$

$$= \iiint_{-\infty}^{\infty} (\boldsymbol{\nabla} \times \boldsymbol{H}) \times \boldsymbol{B} dv - \varepsilon_0 \iiint_{-\infty}^{\infty} \boldsymbol{E} \times (\boldsymbol{\nabla} \times \boldsymbol{E}) dv$$

$$= \mu_0^{-1} \iiint_{-\infty}^{\infty} (\boldsymbol{\nabla} \times \boldsymbol{B} - \boldsymbol{\nabla} \times \boldsymbol{M}) \times \boldsymbol{B} dv - \varepsilon_0 \iiint_{-\infty}^{\infty} \boldsymbol{E} \times (\boldsymbol{\nabla} \times \boldsymbol{E}) dv$$

$$= -\iiint_{-\infty}^{\infty} [\mu_0^{-1} \boldsymbol{\nabla} \times \boldsymbol{M}] \times \boldsymbol{B} dv - \mu_0^{-1} \iiint_{-\infty}^{\infty} \boldsymbol{B} \times (\boldsymbol{\nabla} \times \boldsymbol{B}) dv - \varepsilon_0 \iiint_{-\infty}^{\infty} \boldsymbol{E} \times (\boldsymbol{\nabla} \times \boldsymbol{E}) dv$$

$$= -\boldsymbol{F}_L^{(\text{magnet})}(t) - \iiint_{-\infty}^{\infty} [\boldsymbol{\nabla}(½ \mu_0^{-1} \boldsymbol{B} \cdot \boldsymbol{B} + ½ \varepsilon_0 \boldsymbol{E} \cdot \boldsymbol{E}) - \mu_0^{-1} (\boldsymbol{B} \cdot \boldsymbol{\nabla}) \boldsymbol{B} - \varepsilon_0 (\boldsymbol{E} \cdot \boldsymbol{\nabla}) \boldsymbol{E}] dv. \quad (35)$$

In the above equation, the integral of the gradient vanishes, because both $\boldsymbol{E}$ and $\boldsymbol{B}$ vanish at infinity. As for the remaining terms we have

$$(\boldsymbol{B} \cdot \boldsymbol{\nabla}) \boldsymbol{B} = \partial_x (B_x \boldsymbol{B}) + \partial_y \left( B_y \boldsymbol{B} \right) + \partial_z (B_z \boldsymbol{B}) - (\underbrace{\boldsymbol{\nabla} \cdot \boldsymbol{B}}_{0}) \boldsymbol{B}. \quad (36)$$

Similarly,

$$(\boldsymbol{E}\cdot\boldsymbol{\nabla})\boldsymbol{E}=\partial_x(E_x\boldsymbol{E})+\partial_y\big(E_y\boldsymbol{E}\big)+\partial_z(E_z\boldsymbol{E})-(\boldsymbol{\nabla}\cdot\boldsymbol{E})\boldsymbol{E}. \tag{37}$$

In the absence of free charges and polarization, the last term of Eq.(37) vanishes. The remaining terms appearing in the integral on the right-hand-side of Eq.(35) thus integrate to zero. Therefore, the force $\boldsymbol{F}_L^{(\text{magnet})}(t)$ balances the time-rate-of-change of the Livens EM momentum. If the magnet fades sufficiently slowly, the $E$-field induced in its surrounding space will be negligible, and the EM momentum of the system can be ignored. Therefore, the self-force experienced by the magnet will vanish.

In the presence of a static charge distribution $\rho(\boldsymbol{r})$, the last term in Eq.(37) will become $-\rho(\boldsymbol{r})\boldsymbol{E}(\boldsymbol{r})/\varepsilon_0$, whose effect on Eq.(35) is to augment the Lorentz force by $-\iiint\rho(\boldsymbol{r})\boldsymbol{E}(\boldsymbol{r})dv$. The Lorentz force on the electric charge distribution is thus seen to be the same as it were in the Einstein-Laub formalism. However, the (slowly-fading) magnet does not appear to experience any force at all. We will see in Section 3.3 that the so-called "hidden momentum" is responsible for this discrepancy. Once the effects of hidden momentum are introduced into the Lorentz formalism, the predictions of the Lorentz theory will coincide with those of Einstein and Laub.

**3.2. Conservation of angular momentum as the magnetization fades**. Continuing with the system discussed in Section 3.1, we now analyze the Livens EM angular momentum $\boldsymbol{\mathcal{L}}_L$ of a piece of magnetic material whose magnetization $\boldsymbol{M}(\boldsymbol{r},t)$ varies arbitrarily with time. We write

$$\begin{aligned}
\partial_t\boldsymbol{\mathcal{L}}_L&=\frac{\partial}{\partial t}\iiint_{-\infty}^{\infty}\boldsymbol{r}\times[\varepsilon_0\boldsymbol{E}(\boldsymbol{r},t)\times\boldsymbol{B}(\boldsymbol{r},t)]dv\\
&=\iiint_{-\infty}^{\infty}\boldsymbol{r}\times\left[\varepsilon_0\frac{\partial\boldsymbol{E}(\boldsymbol{r},t)}{\partial t}\times\boldsymbol{B}(\boldsymbol{r},t)\right]dv+\iiint_{-\infty}^{\infty}\boldsymbol{r}\times\left[\varepsilon_0\boldsymbol{E}(\boldsymbol{r},t)\times\frac{\partial\boldsymbol{B}(\boldsymbol{r},t)}{\partial t}\right]dv\\
&=\iiint_{-\infty}^{\infty}\boldsymbol{r}\times[(\boldsymbol{\nabla}\times\boldsymbol{H})\times\boldsymbol{B}]dv-\varepsilon_0\iiint_{-\infty}^{\infty}\boldsymbol{r}\times[\boldsymbol{E}\times(\boldsymbol{\nabla}\times\boldsymbol{E})]dv\\
&=\mu_0^{-1}\iiint_{-\infty}^{\infty}\boldsymbol{r}\times[(\boldsymbol{\nabla}\times\boldsymbol{B}-\boldsymbol{\nabla}\times\boldsymbol{M})\times\boldsymbol{B}]dv-\varepsilon_0\iiint_{-\infty}^{\infty}\boldsymbol{r}\times[\boldsymbol{E}\times(\boldsymbol{\nabla}\times\boldsymbol{E})]dv\\
&=-\iiint_{-\infty}^{\infty}\boldsymbol{r}\times[(\mu_0^{-1}\boldsymbol{\nabla}\times\boldsymbol{M})\times\boldsymbol{B}]dv\\
&\quad\;-\iiint_{-\infty}^{\infty}\boldsymbol{r}\times[\mu_0^{-1}\boldsymbol{B}\times(\boldsymbol{\nabla}\times\boldsymbol{B})+\varepsilon_0\boldsymbol{E}\times(\boldsymbol{\nabla}\times\boldsymbol{E})]dv\\
&=-\boldsymbol{T}_L^{(\text{magnet})}(t)-\iiint_{-\infty}^{\infty}\boldsymbol{r}\times[\boldsymbol{\nabla}(\tfrac{1}{2}\mu_0^{-1}\boldsymbol{B}\cdot\boldsymbol{B}+\tfrac{1}{2}\varepsilon_0\boldsymbol{E}\cdot\boldsymbol{E})-\mu_0^{-1}(\boldsymbol{B}\cdot\boldsymbol{\nabla})\boldsymbol{B}-\varepsilon_0(\boldsymbol{E}\cdot\boldsymbol{\nabla})\boldsymbol{E}]dv.
\end{aligned}\tag{38}$$

The integrand on the right-hand-side of Eq.(38) may now be simplified as follows:

$$\begin{aligned}
\boldsymbol{r}\times\boldsymbol{\nabla}(\tfrac{1}{2}\mu_0^{-1}\boldsymbol{B}\cdot\boldsymbol{B}+\tfrac{1}{2}\varepsilon_0\boldsymbol{E}\cdot\boldsymbol{E})&=(\tfrac{1}{2}\mu_0^{-1}\boldsymbol{B}\cdot\boldsymbol{B}+\tfrac{1}{2}\varepsilon_0\boldsymbol{E}\cdot\boldsymbol{E})(\cancelto{0}{\boldsymbol{\nabla}\times\boldsymbol{r}})\\
&\quad-\boldsymbol{\nabla}\times\big[\big(\tfrac{1}{2}\mu_0^{-1}\boldsymbol{B}\cdot\boldsymbol{B}+\tfrac{1}{2}\varepsilon_0\boldsymbol{E}\cdot\boldsymbol{E}\big)\boldsymbol{r}\big].
\end{aligned}\tag{39}$$

$$\begin{aligned}
\boldsymbol{r}\times(\boldsymbol{B}\cdot\boldsymbol{\nabla})\boldsymbol{B}&=\boldsymbol{r}\times\big(B_x\partial_x\boldsymbol{B}+B_y\partial_y\boldsymbol{B}+B_z\partial_z\boldsymbol{B}\big)\\
&=\partial_x(B_x\boldsymbol{r}\times\boldsymbol{B})+\partial_y\big(B_y\boldsymbol{r}\times\boldsymbol{B}\big)+\partial_z(B_z\boldsymbol{r}\times\boldsymbol{B})-\cancelto{0}{\boldsymbol{B}\times\boldsymbol{B}}-(\cancelto{0}{\boldsymbol{\nabla}\cdot\boldsymbol{B}})(\boldsymbol{r}\times\boldsymbol{B}).
\end{aligned}\tag{40}$$

$$\boldsymbol{r}\times(\boldsymbol{E}\cdot\boldsymbol{\nabla})\boldsymbol{E}=\partial_x(E_x\boldsymbol{r}\times\boldsymbol{E})+\partial_y\big(E_y\boldsymbol{r}\times\boldsymbol{E}\big)+\partial_z(E_z\boldsymbol{r}\times\boldsymbol{E})-\cancelto{0}{\boldsymbol{E}\times\boldsymbol{E}}-(\boldsymbol{\nabla}\cdot\boldsymbol{E})(\boldsymbol{r}\times\boldsymbol{E}). \tag{41}$$

In the absence of external charges and polarization, $\boldsymbol{\nabla}\cdot\boldsymbol{E}=0$, in which case the integral on the right-hand-side of Eq.(38) vanishes. We thus find that the time-rate-of-change of the angular momentum $\boldsymbol{\mathcal{L}}_L(t)$ of the field is equal and opposite to the self-torque $\boldsymbol{T}_L^{(\text{magnet})}(t)$ experienced by the magnet. Once again, if the magnetization fades away slowly, the induced $E$-field in the surrounding space will be negligible, leading to the conclusion that no EM angular momentum would appear in the system and that, therefore, no self-torque will be experienced by the magnet.

In the presence of a static charge distribution $\rho(\boldsymbol{r})$, the term $\boldsymbol{\nabla}\cdot\boldsymbol{E}$ in Eq.(41) will become $\rho(\boldsymbol{r})/\varepsilon_0$, whose contribution to Eq.(38) augments the Lorentz torque by $-\iiint \boldsymbol{r}\times\rho(\boldsymbol{r})\boldsymbol{E}(\boldsymbol{r})dv$. The Lorentz torque acting on electric charges is thus seen to be the same as that obtained in the Einstein-Laub formalism. Once again, the (slowly-fading) magnet does not appear to experience any torque at all. As before, hidden momentum is responsible for this discrepancy, which will be clarified in Section 3.3. With hidden momentum brought into the Lorentz formalism, the predictions of the Lorentz theory will coincide with those of Einstein and Laub.

**3.3. Force, torque, hidden momentum, and hidden angular momentum**. For a static magnet having magnetization profile $\boldsymbol{M}(\boldsymbol{r})$, in the presence of a static $E$-field produced by $\rho(\boldsymbol{r})$ via the associated scalar potential $\psi(\boldsymbol{r})$, we have

$$\begin{aligned}\boldsymbol{\mathcal{P}}_L &= \iiint_{-\infty}^{\infty}\varepsilon_0\boldsymbol{E}(\boldsymbol{r})\times\boldsymbol{B}(\boldsymbol{r})dv\\ &= \iiint_{-\infty}^{\infty}\varepsilon_0\boldsymbol{E}(\boldsymbol{r})\times\boldsymbol{M}(\boldsymbol{r})dv - c^{-2}\iiint_{-\infty}^{\infty}[\boldsymbol{\nabla}\psi(\boldsymbol{r})]\times\boldsymbol{H}(\boldsymbol{r})dv\\ &= \iiint_{-\infty}^{\infty}\varepsilon_0\boldsymbol{E}(\boldsymbol{r})\times\boldsymbol{M}(\boldsymbol{r})dv - c^{-2}\iiint_{-\infty}^{\infty}[\boldsymbol{\nabla}\times(\psi\boldsymbol{H}) - \psi\cancelto{0}{\boldsymbol{\nabla}\times\boldsymbol{H}}]dv.\end{aligned} \tag{42}$$

The second integral on the right-hand-side of Eq.(42) vanishes, and the EM (Livens) momentum becomes equal to the integral of $\varepsilon_0\boldsymbol{E}\times\boldsymbol{M}$ over the volume of the magnet. In general, this momentum is said to be equal in magnitude and opposite in direction to the "hidden" mechanical momentum of the magnet in the presence of an $E$-field.[8,17-39] The sum of the EM and hidden momenta thus vanishes, as expected from a static system whose center of mass-energy is believed to be stationary. The hidden momentum-density is thus expressed as

$$\boldsymbol{\wp}_{\text{hidden}}(\boldsymbol{r}) = \varepsilon_0\boldsymbol{M}(\boldsymbol{r})\times\boldsymbol{E}(\boldsymbol{r}). \tag{43}$$

As explained in conjunction with Eq.(35), in the Lorentz formalism, the force of the induced $E$-field on $\rho(\boldsymbol{r})$ is *not* compensated by an equal and opposite force exerted by $\rho(\boldsymbol{r})$ on the magnet. Instead, the latter force is said to be produced by the time-rate-of-change of the hidden momentum, namely,

$$\boldsymbol{F}^{(\text{magnet})}(t) = -\frac{\partial}{\partial t}\iiint_{-\infty}^{\infty}\boldsymbol{\wp}_{\text{hidden}}(\boldsymbol{r},t)dv = \iiint_{-\infty}^{\infty}[-\partial_t\boldsymbol{M}(\boldsymbol{r},t)\times\varepsilon_0\boldsymbol{E}(\boldsymbol{r})]dv. \tag{44}$$

The minus sign in the above equation indicates that the hidden momentum does not merely disappear; rather it is transferred to the magnet as "overt" mechanical momentum. The right-hand-side of Eq.(44) is now seen to be identical to the Einstein-Laub force obtained in the analysis of Section 2.1; see, in particular, the discussion surrounding Eq.(20).

In similar fashion, the hidden angular momentum in the Lorentz formalism is believed to have the density $\boldsymbol{\ell}_{\text{hidden}}(\boldsymbol{r},t) = \boldsymbol{r}\times\boldsymbol{\wp}_{\text{hidden}}(\boldsymbol{r},t)$. Any change in the angular momentum hidden inside a magnet will give rise to a torque on the magnet in accordance with the formula

$$\boldsymbol{T}^{(\text{magnet})}(t) = -\frac{\partial}{\partial t}\iiint_{-\infty}^{\infty}\boldsymbol{r}\times\boldsymbol{\wp}_{\text{hidden}}(\boldsymbol{r},t)dv = \iiint_{-\infty}^{\infty}\boldsymbol{r}\times[-\partial_t\boldsymbol{M}(\boldsymbol{r},t)\times\varepsilon_0\boldsymbol{E}(\boldsymbol{r})]dv. \tag{45}$$

The minus sign in Eq.(45) indicates that the hidden angular momentum does not merely disappear; rather it is transferred to the magnet as "overt" mechanical angular momentum. The right-hand-side of Eq.(45) is now seen to be identical to the Einstein-Laub torque obtained in the analysis of Section 2.3; see, in particular, the discussion surrounding Eq.(30).

In conclusion, the Einstein-Laub formalism treats the questions of electromagnetic force, torque, momentum, and angular momentum in straightforward and transparent manner. We have discussed other advantages of the method of Einstein and Laub in a recent publication.[48]

## Endnote

In the literature [9,12,16], Abraham's stress tensor is usually written as a symmetrized version of Minkowski's tensor, that is,

$$\overleftrightarrow{\mathcal{T}}_A(\boldsymbol{r},t) = ½\left[(\boldsymbol{D}\cdot\boldsymbol{E}+\boldsymbol{B}\cdot\boldsymbol{H})\overleftrightarrow{\mathrm{I}} - (\boldsymbol{DE}+\boldsymbol{ED}) - (\boldsymbol{BH}+\boldsymbol{HB})\right].$$

Abraham's concerns, as well of those of his followers, were primarily with linear, isotropic media, namely, media for which $\boldsymbol{D}=\varepsilon_o\varepsilon\boldsymbol{E}$ and $\boldsymbol{B}=\mu_o\mu\boldsymbol{H}$. In such cases, since the stress tensor of Minkowski, given by Eq.(6), is already symmetric, the above act of symmetrization does not modify the tensor. In Abraham's own 1909 paper [11], the stress tensor is written explicitly only twice, in Eqs.(Va) and (56), and in both instances it is identical to Minkowski's (asymmetric) tensor. At several points in his papers [11], Abraham mentions the symmetry of his tensor, but it appears that he has the special case of linear, isotropic media in mind. The special symmetry that Abraham introduced into Minkowski's theory is, of course, that between the energy flow rate, $\boldsymbol{E}\times\boldsymbol{H}$, and the electromagnetic momentum density, $\boldsymbol{E}\times\boldsymbol{H}/c^2$, which reside, respectively, in the fourth column and the fourth row of the stress-energy tensor. Be it as it may, the expression of the Minkowski stress tensor in Eq.(6) is also being taken here as the asymmetric version of Abraham's (3×3) stress tensor.